\documentclass[twocolumn,showpacs,preprintnumbers,amsmath,amssymb,floatfix]{revtex4-1}

\usepackage{bm}
\usepackage{dcolumn}
\usepackage{graphicx}
\usepackage{color}
\usepackage{epsf}
\usepackage{epsfig}

\newlength{\myhspace}
\newcommand{\hs}{\hspace{\myhspace}}
\begin{document}

\title{Gravitomagnetic Accelerators}

\author{C. Chicone}

\affiliation{Department of Mathematics and Department of Physics and Astronomy, University of Missouri, Columbia,
Missouri 65211, USA }

\author{B. Mashhoon}

\affiliation{Department of Physics and Astronomy, University of Missouri,
Columbia, Missouri 65211, USA}

\begin{abstract}We study a simple class of time-dependent rotating Ricci-flat
cylindrically symmetric spacetime manifolds whose geodesics admit
gravitomagnetic jets. The helical paths of free test particles in these jets
up and down parallel to the rotation axis are analogous to those of charged
particles in a magnetic field. The jets are attractors. The jet speed
asymptotically approaches the speed of light. In effect, such source-free
spacetime regions act as ``gravitomagnetic accelerators.''

\end{abstract}

\maketitle
The purpose of this Letter is to point out that general relativity in principle permits the existence of dynamic source-free spacetime regions in which the speed of free test particles can rapidly increase so as to approach the speed of light.

Mechanisms for the acceleration of particles to ultrarelativistic speeds are important for the explanation of high-energy astrophysical phenomena~\cite{1,2}. Most of the known acceleration processes involve charged particles and are electromagnetic in origin~\cite{3}. Gravitational mechanisms involving arbitrary test particles have been the subject of recent investigations~\cite{4,5,6,7,8}. We adopt a different approach here and present a simple explicit solution of Einstein's vacuum field equations for which the geodesic equation has exact analytic solutions that represent gravitomagnetic jets. The jets are attractors and the speed of a free test particle in such a jet asymptotically approaches the speed of light.

Consider a warped product spacetime with a metric of the form
\begin{equation}\label{eq:1}
ds^2=e^t d\Sigma^2+e^{-t} dz^2,
\end{equation}
where $d\Sigma^2$ is a 3D stationary metric given by
\begin{equation}\label{eq:2}
d\Sigma^2=\frac{X_r}{X}(-X^2 dt^2+\frac{1}{r^3} dr^2) +\frac{1}{r}(-Xdt +d\phi)^2
\end{equation}
and we use units such that $c=1$ throughout.
Here $X_r=dX/dr$ and $X$ is the solution of the differential equation
\begin{equation}\label{eq:fode}
r^2 X^2 X_{rr}+X_r=0.
\end{equation}
The spacetime metric~\eqref{eq:1} has signature $+2$ and  is Ricci-flat, namely, $R_{\mu\nu}=0$~\cite{9}. The cylindrical coordinates $x^\mu=(t,r,\phi,z)$ are dimensionless. To transform them to physical coordinates $\tilde x^\mu=(\lambda' t,\lambda^{-1} r,\phi, \lambda z)$, we need arbitrary lengthscales $\lambda'$ and $\lambda$; then, $\tilde X=\lambda^{1/2} X$ and $\tilde s=\lambda s$. The conformal factor $\exp(t)$, the warping function $\exp(-t)$ and the gravitomagnetic potential $X$ lead to interesting aspects of the  gravitational field under consideration here.

The cylindrical coordinates in the spacetime domain of interest must be admissible. It follows from the Lichnerowicz conditions~\cite{10} that the principal minors of the metric tensor and its inverse must be negative. Hence, $X X_r>0$ and $XQ>0$, where $Q=r X_r-X$. The first of these conditions is satisfied if $X^2$ is a monotonically \emph{increasing} function of $r$. It turns out that Eq.~\eqref{eq:fode} is a special case of the generalized Emden-Fowler equation. Of the two known explicit solutions of this equation, $X=\mathrm{constant}$ is not acceptable as metric~\eqref{eq:1} would then degenerate into a 3D metric and $X=\pm (3r/2)^{-1/2}$ is such that $X^2$ is monotonically \emph{decreasing} with $r$. In fact, with $r\ge 0$ taken as radial coordinate, the solutions of Eq.~\eqref{eq:fode} other than $X=\mathrm{constant}$ fall into two classes: either $X^2$ is monotonically increasing or decreasing. In this work we focus on the former class, since the latter class involves rotating gravitational waves and was investigated in detail in Ref.~\cite{9}.

\begin{figure}[ht]
\begin{center}
\includegraphics[width=35mm]{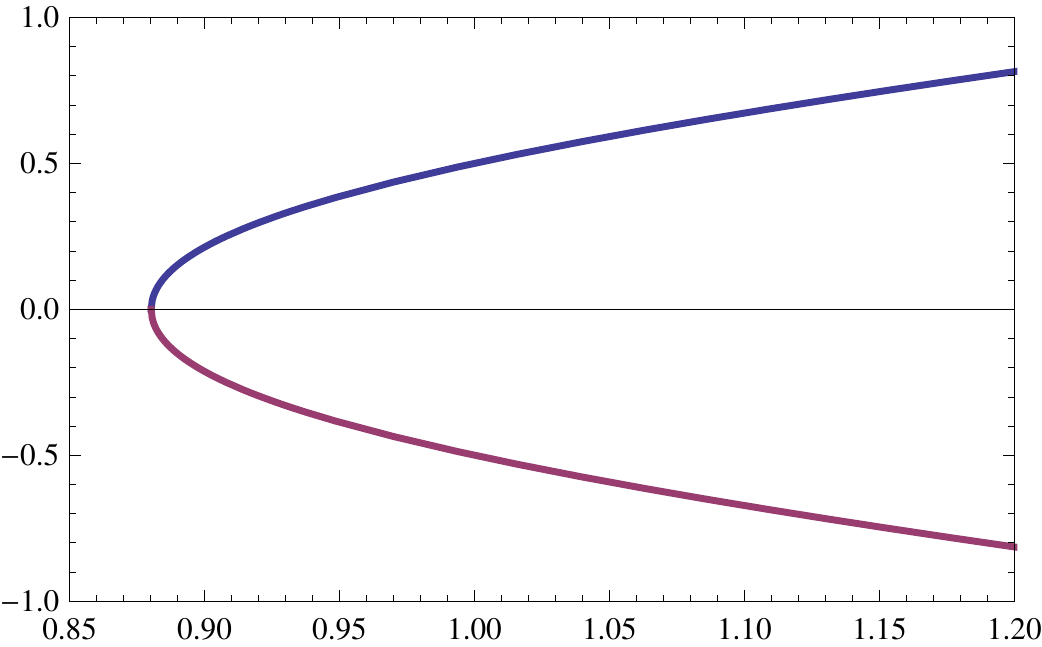}\quad\includegraphics[width=35mm]{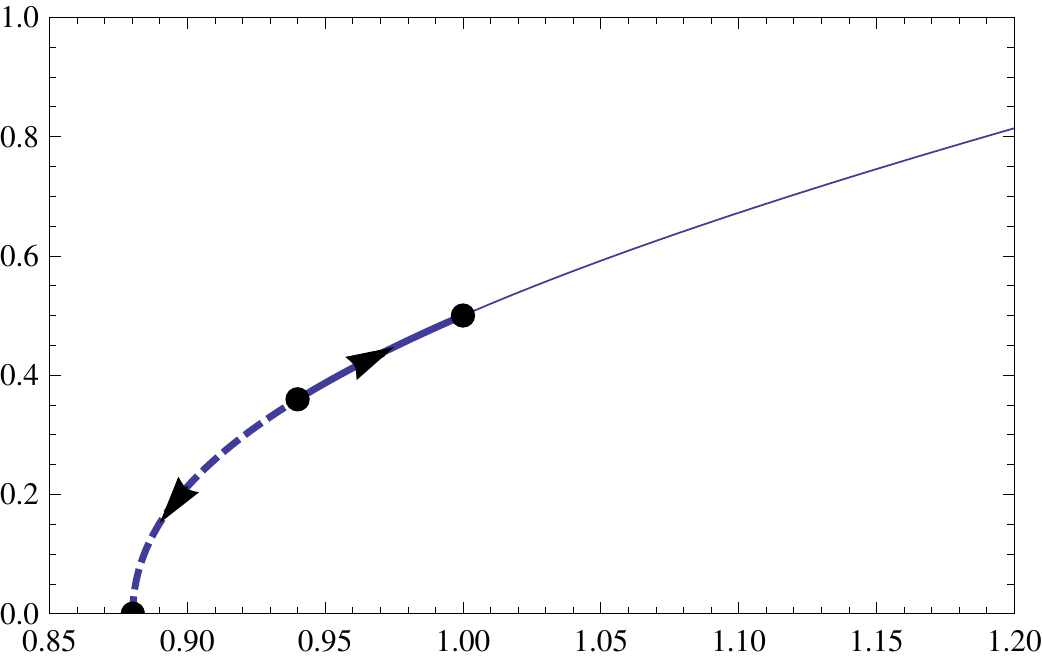}
\end{center}
\caption{Left panel: Plot of the solution $\pm X$ versus $r$ for the differential Eq.~\eqref{eq:fode}. The initial data  are $X(1)=1/2$ and $X_r(1)=2$. The left endpoint occurs at $r_b\approx 0.88$.  This is an admissible solution of Eq.~\eqref{eq:fode} that allows a jet at $r_J=1$. Right panel: Starting from rest at $r_0 = 0.94$, halfway between $r_b$ and $r_J$, the geodesic equation is integrated for a free test particle as in Fig. ~\ref{fig:2}. For $0 \le C_\phi\le 1.65$, the particle is attracted to the jet at  $ r_J$  (right arrow), while
for $C_\phi \ge  1.7$ the particle moves to $r_b$ and exits $\mathcal{S}$ (left arrow).
Similarly, for $ -5 \le C_\phi \le 0$  the particle is attracted to the jet, while
for $C_\phi \le -6$  the particle exits via $r_b$.
\label{fig:1}}
\end{figure}

The solutions of Eq.~\eqref{eq:fode} with $X^2$ monotonically increasing all have the general form depicted in Figure~\ref{fig:1}. If $X$ is a solution of Eq.~\eqref{eq:fode}, then so is $-X$; henceforth, we consider only the $X\ge 0$ branch. It is possible to show that there is an open set of initial conditions corresponding to \emph{admissible} solutions $X$; each exists on a radial interval $(r_b,\infty)$, where $r_b>0$ with limiting values $X(r_b)=0$ and $X_r(r_b)=\infty$ such that $XX_r=r_b^{-2}$ at $r_b$. Furthermore,  $X^2$ increases, $X_r^2$ decreases and $X^2$ approaches infinity as $r\to \infty$. Indeed, $r=\infty$ at  the axis of cylindrical symmetry and $r=r_b$ at the exterior boundary cylinder. The physical region of interest $\mathcal{S}$ is the open hollow expanding cylindrical domain with inner boundary around the axis $r=\infty$ and outer boundary at the null hypersurface $r=r_b$. The cylindrical coordinates are not admissible at these boundaries, since $g^{tt}$ vanishes at $r=\infty$ and $(-g)^{-1/2}$ vanishes at $r=r_b$.

With admissible cylindrical coordinates in $\mathcal{S}$,  we have an algebraically general Petrov type I solution of Einstein's source-free equations with two commuting spacelike Killing vector fields $\partial_\phi$ and $\partial_z$ associated with cylindrical symmetry.  The two-parameter isometry group is not orthogonally transitive. While $\partial_ z$ is hypersurface orthogonal, $\partial_\phi$  is not; moreover, the other two coordinates $(t,r)$ can be invariantly defined via the magnitudes of these Killing vector fields. The $t=\mathrm{constant}$ hypersurfaces are spacelike, while $r=\mathrm{constant}$ hypersurfaces are timelike in the physical domain $(r_b,\infty)$. Near the axis ($r\to \infty$), 
\begin{equation}\label{eq:4}
X(r)\sim a r-\alpha-\frac{1}{6 a r^2}-\frac{\alpha}{6 a^2 r^3}+O(\frac{1}{r^4}),
\end{equation}
where $a>0$ and $\alpha>0$ are constants. Near the boundary ($r\to r_b^+$), 
\begin{equation}\label{eq:5}
X(r)= \frac{2^{1/2}}{r_b} (r-r_b)^{1/2}+\mathcal{A}(r-r_b)+O((r-r_b)^{3/2}),
\end{equation}
where $\mathcal{A}$ is a constant.  We note that solutions of Einstein's gravitational field equations with cylindrical symmetry have been the subject of numerous investigations (see~\cite{11,12,13,14,15} and the references cited therein).

The proper radial distance from the axis to an event with $r=r_0$ is given by 
\begin{equation}\label{eq:6}
e^{t/2}  \int_{r_0}^\infty \Big(\frac{X_r}{r^3 X}\Big)^{1/2}\,dr.
\end{equation}
At a finite instant of time $t$, the proper radial distance from the axis to the boundary is finite. It is possible to show that the condition of elementary flatness is not satisfied near the axis~\cite{16}. Within the spacetime region of interest $\mathcal{S}$, however, there are no curvature singularities. To illustrate this fact, we note that there are four algebraically independent scalar polynomial curvature invariants in a Ricci-flat spacetime that can be represented as~\cite{12}
\begin{align}
\label{eq:13a} \mathcal{I}_1&=R_{\mu\nu\rho\sigma}R^{\mu\nu\rho\sigma}-iR_{\mu\nu\rho\sigma}R^{*\mu\nu\rho\sigma}, \\
\label{eq:14a} \mathcal{I}_2&=R_{\mu\nu\rho\sigma}R^{\rho\sigma\alpha\beta}R_{\alpha\beta}^{\hspace{.2in}\mu\nu}+iR_{\mu\nu\rho\sigma}R^{\rho\sigma\alpha\beta}{R^*}_{\alpha\beta}^{\hspace{.2in}\mu\nu}.
\end{align}
These turn out to be real in this case and can be expressed as 
\begin{align}
\label{eq:15a} \mathcal{I}_1&=-\frac{e^{-2t}}{r X^4X_r^2}(F^2-r X^2 (F+4)), \\
\label{eq:16a} \mathcal{I}_2&=-\frac{3e^{-3t}}{4r X^5X_r^3} F (F+2r X^2),
\end{align}
where $F=r^2 X X_r-1$. These invariants are well behaved in the interior of the physical region $(r_b, \infty)$ and have proper limits at the boundaries. Let us note that $\mathcal{I}_1$ and $\mathcal{I}_2$ both diverge in the infinite past ($t\to -\infty$),  a situation that is consistent with the emergence of the universe from a singular state as in the standard cosmological models.

The spacetime domain $\mathcal{S}$ rotates about the axis of cylindrical symmetry. To elucidate the gravitomagnetic aspects of $\mathcal{S}$,  imagine the class of fundamental observers in $\mathcal{S}$. They are spatially at rest by definition,  with spatial frames that we can choose to be along the natural directions of the cylindrical coordinate system. A unit gyro carried by these observers precesses about the $z$ axis with frequency $(2 X Q)^{-1}$, indicating the presence of a gravitomagnetic field parallel to the $z$ axis~\cite{17}. There is experimental evidence for gravitomagnetism; indeed, GP-B has recently measured the exterior gravitomagnetic field of the Earth~\cite{18}.

We now turn to the motion of free test particles in this gravitational field.  Let $u^\mu=dx^\mu/d\tau$ be the four-velocity vector of the test particle, where $\tau$ is the proper time along its world line. The components of $u^\mu$ along the Killing vector fields will be constants of geodesic motion. Thus $C_z=g_{z\alpha}u^\alpha$ is the constant linear momentum (per unit mass) of the test particle parallel to the $z$ axis, while $C_\phi=g_{\phi\alpha}u^\alpha$ is the constant angular momentum (per unit mass) of the test particle about the $z$ axis. That is,
\begin{equation}\label{eq:noether}
\frac{dz}{d\tau}=C_z e^t,\qquad \frac{d\phi}{d\tau}- X \frac{dt}{d\tau}= C_\phi r e^{-t}.
\end{equation}
It then follows from $u^\alpha u_\alpha=-1$ that $dt/d\tau=V/X$, where
\begin{equation}\label{eq:12}
V=\Big[\frac{1}{r^3} U^2 +\frac{X}{X_r} (e^{-t}+C_z^2+C_\phi^2 r e^{-2t})\Big]^{1/2}.
\end{equation}
Here $U=dr/d\tau$ and we have assumed that $t$ monotonically increases with $\tau$ along the geodesic world line. Thus the geodesic equation reduces in this way to the radial equation given by
\begin{equation}\label{eq:13}
\frac{dU}{d\tau}=-\Gamma^r_{\alpha\beta}u^\alpha u^\beta.
\end{equation}
In practice, Eq.~\eqref{eq:fode} must also be solved simultaneously with respect to the particle's proper time. 

\begin{figure}[ht]
\begin{center}
\includegraphics[width=25mm]{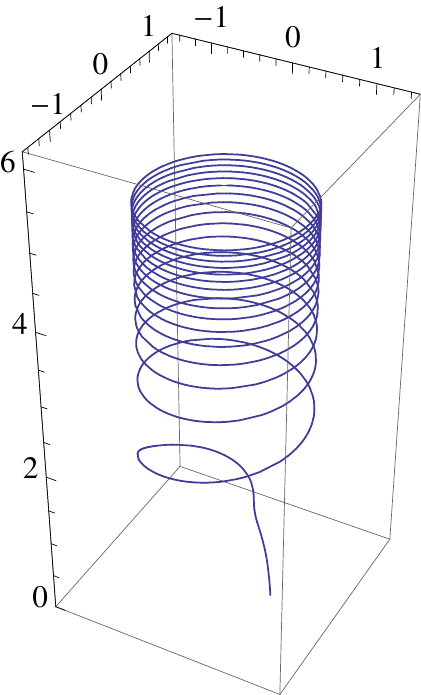}\quad \includegraphics[width=25mm]{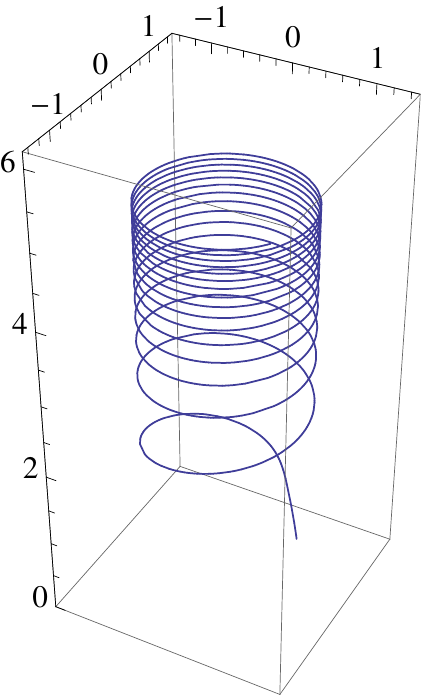}
\end{center}
\caption{The result of integration of the geodesic equation for timelike geodesics attracted to the jet associated with $X$ given in Fig.~\ref{fig:1}.  The initial data at $\tau_0=0$ are  
$t_0 = 0$,
$r_0 = 0.94$, $U_0 = 0$,
$\phi_0 = 0$,
$z_0 = 0$,
$X(r_0) \approx 0.36$ and 
$X_r(r_0) \approx 2.84$. In the left-hand plot the parameters are $C_\phi = -5$ and
$C_z = 1$. The right-hand plot  is for  $C_\phi = 1.65$ and  $C_z=1$.  The coordinates are $(x,y,\hat z)$, where $x=\rho\cos\phi$, $y=\rho\sin\phi$ and $\rho=r^{-1/2}$; moreover, $\hat z=\ln|\ln z|$. We use $\hat z$ instead of $z$ for the sake of clarity.
\label{fig:2}}
\end{figure}
In analogy with the motion of a charged particle in a magnetic field, we look for geodesics that are confined to a cylinder of fixed radius $r\in (r_b,\infty)$. Thus we set $U=0$ and $dU/d\tau=0$. Equations~\eqref{eq:noether} and~\eqref{eq:12} yield
\begin{equation}\label{eq:14}
\frac{dt}{d\phi}=\frac{V}{X(V+C_\phi re^{-t})},
\end{equation}
while it follows from Eq.~\eqref{eq:13} that 
\begin{equation}\label{eq:15}
\frac{dt}{d\phi}=-\frac{1}{Q\pm (X_r/X)^{1/2}}.
\end{equation}
These are compatible, the latter with the lower sign only,   if and only if $C_\phi=0$ and  there is a radial coordinate $r_{J}\in (r_b,\infty)$ such that  $r^2 X X_r=1$ for $r=r_J$. In this  case, the geodesic equation can be solved explicitly for a special  three-parameter set of geodesics parameterized by the initial values of $(t,\phi,z)$ such that  $r$ remains constant at $r_J$ . The   geodesics in this class exhibit helical motions up ($C_z>0$) and down ($C_z<0$) on the cylinder $r=r_J$ parallel to the $z$ axis except for the measure zero subset  ($C_z=0$) whose motions remain bounded on circular orbits about the $z$ axis.  It can be shown that for a given $X$, either there is a unique $r_J$,  $F(r_J)=0$,  or there is none; moreover,  there is a non-empty open subset of admissible solutions of Eq.~\eqref{eq:fode}, corresponding to  $\mathcal{A}<0$ in Eq.~\eqref{eq:5},  that allow these special geodesics.  For a  solution of Eq.~\eqref{eq:fode}  in this class,  we call the set of special geodesics a \emph{gravitomagnetic jet}.  

Numerical experiments reveal that gravitomagnetic jets are attractors, see Figs.~\ref{fig:1} and~\ref{fig:2}. That is,  the union of  special geodesics with a given $X$ and $r_J$ is a non-compact connected invariant manifold that attracts all nearby geodesics. Figure~\ref{fig:2} highlights the helical motion of a gravitomagnetic jet. We note that these helical motions up and down within a double-jet configuration have the \emph{same} orientation;  the helical sense is positive in our case due to our choice of the $X\ge 0$ branch.  For recent studies of helical motions in astrophysical jets, see~\cite{19}.

Null geodesics can be treated similarly in $\mathcal{S}$ and it turns out that they have special helical solutions confined to $r=r_J$ just as in the case of  timelike geodesics. Moreover, for $t\to \infty$, special timelike geodesics go over to special null geodesics as a simple consequence of the dynamical equations of motion. 

To explore this important aspect of gravitomagnetic jets further, we consider the speed of jets with respect to the fundamental observers. These are endowed with an orthonormal tetrad  $\lambda^\mu_{\hs (\alpha)}$ such that in $(t, r,\phi, z)$ coordinates
 \begin{equation}\label{eq:16}
\lambda^\mu_{\hs (t)}=((-g_{tt})^{-1/2},0,0,0)
\end{equation}
and we can choose the spatial frame $\lambda^\mu_{\hs(i)}$ to be along the standard cylindrical coordinate axes. Then $u^\mu=u^{(\alpha)}\lambda^\mu_{\hs (\alpha)}$, where $u^{(\alpha)}=\gamma(1,\mathbf{v})$ is the jet four-velocity measured by the fundamental observers. Hence, $\gamma=-u_\mu\lambda^\mu_{\hs (t)}$ and a straightforward calculation reveals that
 \begin{equation}\label{eq:17}
\gamma=\gamma_0(1+C_z^2 e^t)^{1/2}.
\end{equation}
Here $\gamma_0$ is the Lorentz factor corresponding to circular motion with $C_z=0$; that is, $\gamma_0=(1-\beta_0^2)^{-1/2}$, where $\beta_0^2=r_J X^2(r_J)$. For $C_z\ne 0$ and $t\to \infty$, $\gamma$ diverges exponentially with time; that is, the jet speed rapidly approaches the speed of light. The gravitational influence of the test particle on the
spacetime geometry has been neglected in our work;  clearly, this
approximation eventually breaks down for the gravitomagnetic jets.

Equation~\eqref{eq:fode} is invariant under the scale transformation $(r,X)\mapsto (\hat r, \hat X)$, where $\hat r=\sigma r$ and $\hat X=\sigma^{-1/2} X$ for $\sigma\in (0,\infty)$. This scale invariance can be used to reduce Eq.~\eqref{eq:fode} to a first order system of autonomous equations. For instance, let $\theta=-\ln X$ and consider scale-invariant  variables $p$ and $q$, where $p(\theta)=X/(r X_r)$ and $q(\theta)=(F+1)^{-1}$; then, Eq.~\eqref{eq:fode} reduces to the Lotka-Volterra system 
\begin{equation}\label{eq:18}
\frac{dp}{d\theta}=p(p-q-1), \qquad \frac{dq}{d\theta}= q(2 p-q+1).
\end{equation}
A solution $X$ is admissible when $0<p<1$ for all $\theta$; similarly, an admissible solution allows jets when $q(\theta)=1$ for some $\theta>0$, where $\theta=-\infty$ at the symmetry axis and $\theta=\infty$ at the outer boundary. A detailed investigation of system~\eqref{eq:18}---for admissible solutions that allow jets---reveals that $p/q=r X^2$ for $q=1$ lies in the interval $(0,j)$, where $j\approx 0.4$. This means that $\beta_0$, the speed of the free test particle on a circle of radius $r_J$, is such that $\beta_0\in (0, j^{1/2})$, where $j^{1/2}\approx 0.63$. 

It follows from these results that free test particles in a gravitomagnetic jet can in principle start out with speeds near zero, but they then inevitably undergo rapid  ``acceleration"  to almost the speed of light. Similarly, one can study the motion of other geodesics in $\mathcal{S}$ with respect to the fundamental observers. In the simple numerical experiments whose results are  presented in Figs.~\ref{fig:1}  and~\ref{fig:2},  we followed the variation of the Lorentz factor from the initial point to late times. There is indeed a vast difference between the geodesics that leave $\mathcal{S}$ via
$r_b$  and those that are attracted to the jet---see the right panel of Fig.~\ref{fig:1}. In the former case, the Lorentz factor initially decreases but then increases as the geodesic exists $\mathcal{S}$, remaining within about an order of magnitude of unity.  In the latter case, the Lorentz factor quickly diverges to infinity---starting from $\approx 3.5$  (for the left-hand plot) and $\approx  2.9$ (for
the right-hand plot) in Fig.~\ref{fig:2}.

The existence of a jet is a scale-invariant property; that is, for every admissible solution with a jet, scaling leads to a one-parameter family of solutions of the same kind, where the jet now occurs at $\sigma r_J$. One can use this property to set $r_J=1$ for every jet solution. Thus Eq.~\eqref{eq:fode} can be integrated with initial conditions that at $r_J=1$, $X(1)=\vartheta^{-1}$ and $X_r(1)=\vartheta$; in this case, $r_b\approx 1-\vartheta^{-2}/2$ for $\vartheta\gg 1$. The solution is admissible once $\beta_0^2=\vartheta^{-2}<j$; that is, $\vartheta>j^{-1/2}\approx 1.6$. For example, in Fig.~\ref{fig:1} we have $\vartheta=2$ and hence $\beta_0=0.5$ for the corresponding jet depicted in Fig.~\ref{fig:2}. A detailed treatment of gravitomagnetic jets is contained in Ref.~\cite{20}.

 The empty annular region of physical interest that allows gravitomagnetic
jets could in principle be joined to two concentric circular cylindrical
regions with matter such that the inner one surrounds the axis ($r =
\infty$) and the outer one has its inner boundary at $r = r_b$.
Our treatment has been greatly simplified by the assumption of
cylindrical symmetry; nevertheless, we hope that in the context of general
relativity similar gravitomagnetic accelerators may emerge under physically
more realistic circumstances. In any case, this appears to be a promising
approach in the search for the origin of high-energy astrophysical jets.  Further recent results in this direction are contained in Ref.~\cite{21}.

The work of CC was supported in part by the NSF grant DMS 0604331.

\end{document}